\documentclass[journal]{IEEEtran}

\ifCLASSOPTIONcompsoc
\else
\fi

\ifCLASSINFOpdf
\else
\fi
\usepackage{graphicx}

\usepackage{multirow}

\begin{document}

\title{A Change Support Model for Distributed Collaborative Work}

\author{\IEEEauthorblockN{Phan Thi Thanh Huyen and Koichiro Ochimizu}\\
\IEEEauthorblockA{School of Information Science\\
Japan Advanced Institute of  Science and Technology (JAIST) \\
Nomi, Ishikawa, Japan\\
Email: \{huyenttp, ochimizu\}@jaist.ac.jp}

}

%\IEEEauthorblockA{\IEEEauthorrefmark{1}Current company: Internet Initiative Japan (IIJ)\\}}


%Telephone: (800) 555--1212, Fax: (888) 555--1212}


\maketitle

% display page numbers in the headings. Start with roman numerals %
\pagestyle{headings}
\setcounter{page}{1}
%\pagenumbering{roman}

\begin{abstract}
Distributed collaborative software development tends to make artifacts and decisions inconsistent and uncertain. 
We try to solve this problem by providing an information repository to reflect the state of works precisely, by managing the states of artifacts/products made through collaborative work, and the states of decisions made through communications.
In this paper, we propose models and a tool to construct the artifact-related part of the information repository, and explain the way to use the repository to resolve inconsistencies caused by concurrent changes of artifacts.
We first show the model and the tool to generate the dependency relationships among UML model elements as content of the information repository. Next, we present the model and the method to generate change support workflows from the information repository. These workflows give us the way to efficiently modify the change-related artifacts for each change request. Finally, we define inconsistency patterns that enable us to be aware of the possibility of inconsistency occurrences. By combining this mechanism with version control systems, we can make changes safely.
Our models and tool are useful in the maintenance phase to perform changes safely and efficiently.

%By using CSWs to implement change requests, in contrast to previous work, change workers can have very comprehensive views of shared software artifacts, of related workers and their previous, current, and future works. Thus, potential inconsistencies can be detected earlier to help workers to implement timely adjustments. CSM is of practical usefulness because it enables change workers to conduct change activities safely and efficiently in a collaborative environment.

\end{abstract} 


%\textbf{\textit{Change Support Workflow; dependency; inconsistency; awareness; UML model element}}
%\textbf{Keywords: \textit{Inconsistency Management; Generation of Dependency Relationship; Change Support Workflow; Awareness of Inconsistency}}

% \fi
\IEEEpeerreviewmaketitle

\section{Introduction}
\begin{figure}[!b]
\centering
\includegraphics[scale = .4]{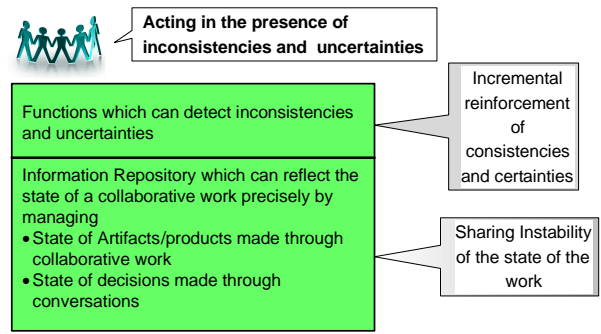}
\caption{An approach to acting in the presence of inconsistencies and uncertainties in a distributed collaborative environment}
\label{fig_generalApproach}
\end{figure}

Collaboration is unavoidable in software development because of the increased scale and complexity of projects. However, collaboration is not an easy task, and bad collaboration contributes to project failure.  One common problem in a collaborative work is different understandings of the state of shared artifacts, interface definitions, and agreements made through communications. These recognition gaps are more serious in a distributed environment, where geographical distribution of workers can cause convergence delay. J. D. Herbsleb reported that ``distributed work items appear to take about two and one-haft times as long to complete as similar items where all the work is collocated" \cite{ref_Empirical_study_speed_communication_JD.Herbsleb}. Therefore, in addition to ordinary software development environments (SDEs), additional support for distributed collaborative work is necessary. In \cite{ref_instability_ochimizu}, \cite{ref_software_process_model_Ochimizu}, we have presented an approach to deal with the instability of the state of distributed collaborative work, which tends to make artifacts and decisions inconsistent and uncertain. This approach proposed technologies for sharing the instability of the state of the distributed collaborative work, and for incremental reinforcement of consistencies and certainties (Fig. \ref{fig_generalApproach}).
%, as shown in Fig. \ref{fig_generalApproach}.

%Regarding the awareness of uncertainty, we have proposed a method for decision
Regarding the uncertainty problem, we have proposed a method for decision management support. In a distributed collaborative environment, email is one of the most popular means of communication among workers. However, in email communications, a discussion topic may involve many emails, and one email may contain many topics. Therefore, it is helpful for the participants in deliberations to grasp the progress and the reasoning of the deliberation for each subject. In \cite{ref_deliberation_thread}, we have presented a model and a tool for extracting deliberation threads from email communications. This tool helps to reduce the gap in understanding collaborative work among workers, and leads them to reach a decision.

%Toward awareness support of inconsistency, this paper will present a Change Support Model (CSM) for distributed collaborative work.  Our CSM is a combination of model-based approach, process support approach, and awareness support approach, which are the main collaboration techniques in software engineering [2]. One unique feature of CSM is the semi-automatic generation of Change Support Workflows (CSWs) that represent activities needed to implement change requests. CSW is generated based on dependency relationships among UML model elements. Change workers will perform change activities by following these workflows. CSM is of practical usefulness because it enables change workers to conduct change activities safely and efficiently in a collaborative environment. Another feature is that software artifacts changed by these workflows are not limited to source codes but UML model elements. 

Regarding the inconsistency problem, this paper presents a Change Support Model (CSM) for distributed collaborative work, in which we propose models and a tool to construct the artifact-related part of the information repository, and we explain the way to use the repository to resolve inconsistencies caused by concurrent changes of artifacts. CSM is a combination of model-based approach, process support approach, and awareness support approach, 
%which are the
the main collaboration techniques in software engineering \cite{ref_collaboration_roadmap}. CSM is useful in the maintenance phase to perform changes safely and efficiently.

The information repository contains UML model elements with dependency relationships. We will show a model and a tool to generate the dependency relationships among the UML model elements. Next, we present a method to generate a Change Support Workflow (CSW) for each change request from the information repository. The CSW gives us a way to modify the related artifacts for each change request efficiently. Change workers will perform change activities for each change request by following the generated CSW.

Concerning the inconsistency detection, previous studies have concentrated on detecting code conflicts by using synchronization function of version control system, or by monitoring the behaviors of developers in different workspaces. However, in these works, conflicts are detected after changes have been finished or are being executed, and the awareness of developers is limited to the program elements being accessed concurrently by others. In CSM, we identify patterns of inconsistency existing among concurrently executing CSWs to detect the possibility of inconsistency occurrences earlier. In addition, by using CSWs, CSM can provide change workers with very comprehensive views of shared artifacts.

%The rest of this paper is organized as follows. Section 2 provides an overview of our approach to build the CSM. The framework of CSM is also given in this section. In Section 3, we present our method to generate Basic Dependency Relationships (BDR) among UML model elements. This is a preparation for the next step, CSW generation based on the generated relationships. Section 4 is divided into two parts.  The first part describes our method to generate CSWs from relationships among artifacts and change requests. The second part handles inconsistencies among CSWs. A prototype of the system is given in Section 5. Section 6 reports related work, and finally, Section 7 concludes the paper and discusses points to future work.
The rest of this paper is organized as follows. Section 2 provides an overview of our approach to building the CSM and introduces its framework. 
%The framework of CSM is also given in this section. 
We present our method and the tool to generate the dependency relationships among UML model elements in Section 3, and develop the method for generating CSWs from these dependency relationships among artifacts in Section 4. Section 5 handles awareness support regarding inconsistencies. 
%An impact analysis tool that implements the BDR generation method is given in Section 6. 
Section 6 reports related work, and finally, Section 7 concludes the paper and discusses future work.

\section{Approach}

% 7Apr Changes are inevitable during software development and after delivery. In a distributed collaborative environment, change implementation is more difficult because software artifacts with very complex dependency relationships are created based on the collaboration of many workers. Also, lacking awareness of concurrent works of workers contributes to inconsistencies and potential inconsistencies on shared artifacts. Solving these problems is our motivation to build the Change Support Model that helps to deal with the instability and inconsistency of shared software artifacts in the distributed collaborative environment. Some introduction of CSM was first given in [4].
\subsection{Overview}
Distributed collaborative software development tends to make artifacts and decisions inconsistent and uncertain. Our general approach to this problem is to construct an information repository that can reflect the state of work precisely, by managing the states of artifacts/products made through collaborative work and the states of decisions made through communications, and to develop functions that can detect inconsistencies and uncertainties (Fig. \ref{fig_generalApproach}). 
In this paper, we apply this approach to the change environment to deal with the inconsistency problem that occurs when several changes of artifacts are made concurrently.

%Changes are inevitable during software development and after delivery. In a distributed collaborative environment, many changes are made concurrently by different workers who may work independently from each other. Because there are always very complex relationships among software artifacts,  
%Because of insufficient information about the activities of others, workers can not prevent inconsistencies by themselves. 
%Because software artifacts with very complex relationships are created based on the collaboration of many workers. Also, lacking awareness of concurrent works of workers contributes to (potential) inconsistencies on shared artifacts. 

\begin{figure}[!t]
\centering
\includegraphics[scale = 0.44]{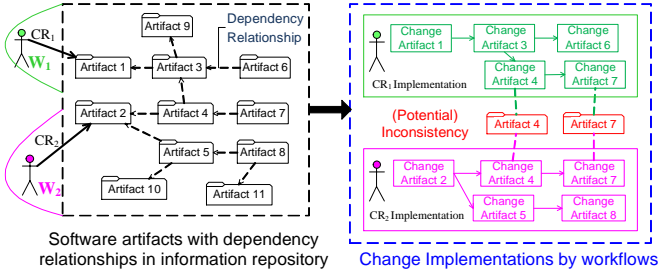}
\caption{Inconsistency in a change environment}
\label{fig_inconsistency_problem}
\end{figure}

Changes are inevitable during software development and after delivery. In a distributed collaborative environment, change implementation is difficult because software artifacts with very complex relationships are created based on the collaboration of many workers. Also, lack of awareness of concurrent work of workers contributes to (potential) inconsistencies on artifacts (See Fig. \ref {fig_inconsistency_problem}). Supporting change workers to work safely and efficiently is the objective of the CSM.

%When a change worker makes a change on an artifacts, artifacts connected to this artifact by dependency relationships may be affected. Repeating this process, we can see that from an initial change on the source artifact, change can spread out to many artifacts that can reach to the source artifact by dependency relationship. 
%When a change worker makes a change on an artifact, the artifacts connected to this artifact by dependency relationships may be affected. Dependency is a relationship between two artifacts in which a change to one artifact may affect to the other. By repeating the process, we can see that a change initiated from a source artifact can spread to many artifacts that in turn can reach the source artifact by dependency relationships. By analyzing dependency relationships among software artifacts, we can estimate impact of a change and steps a change worker should do to implement a change request. Based on these foreseen results, we want to represent explicitly activities needed to implement a change request by using workflows. Suitable generation and management of workflows can help to increase the awareness of work states of other workers and to reduce inconsistencies in the distributed collaborative environment. In the CSM, we define a Change Support Workflow (CSW) as a sequence of activities defined to carry out a change request.  Activities in CSW take care of creating new software artifacts or modifying exiting ones. CSM is responsible for CSW construction and management.
When a change worker makes a change on an artifact, called \textit{change root}, the artifacts connected to this artifact by dependency relationships may be affected. Dependency is a relationship between two artifacts in which a change to one artifact may affect the other. By repeating the process, we can see that a change initiated from a change root can spread to many artifacts, which in turn can reach the change root by dependency relationships. In the example shown in Fig. \ref {fig_inconsistency_problem}, from a change request (CR) on artifact 1 ($CR_1$), Worker 1 ($W_1$) may have to implement changes on artifacts 3, 4, 6, and 7. Similarly, from $CR_2$ on artifact 2, Worker 2 ($W_2$) may have to implement changes on artifacts 2, 4, 5, 7, and 8. Because the two workers work independently, they may not have sufficient information about each other's activities. Therefore, if these change requests are implemented concurrently, inconsistency may happen on shared artifacts 7 and 8. To implement changes efficiently, we will formalize change implementations on related artifacts for each change request in the system by a special workflow, Change Support Workflow (CSW). A CSW is a sequence of activities defined to carry out a change request. Activities in a CSW take care of creating new artifacts or modifying exiting ones. To implement changes safely, a mechanism to handle inconsistency is indispensable in CSM.

% Basic Dependency Relationships (BDRs) belong to inter-relationship group . 

%To increase the awareness of work state of other workers and to reduce inconsistencies in the distributed collaborative environment, we use workflow to represent activities needed to implement a change request. We define a Change Support Workflow (CSW) as a sequence of activities defined to carry out a change request.  Activities in CSW take care of creating new software artifacts or modifying exiting ones. CSM is responsible for CSW construction and management.
\subsubsection{Dependency Generation} To generate a CSW for a change request, we need to identify the impacted artifacts based on dependency relationships among artifacts.

\begin{figure}[!t]
\centering
\includegraphics[scale=0.25]{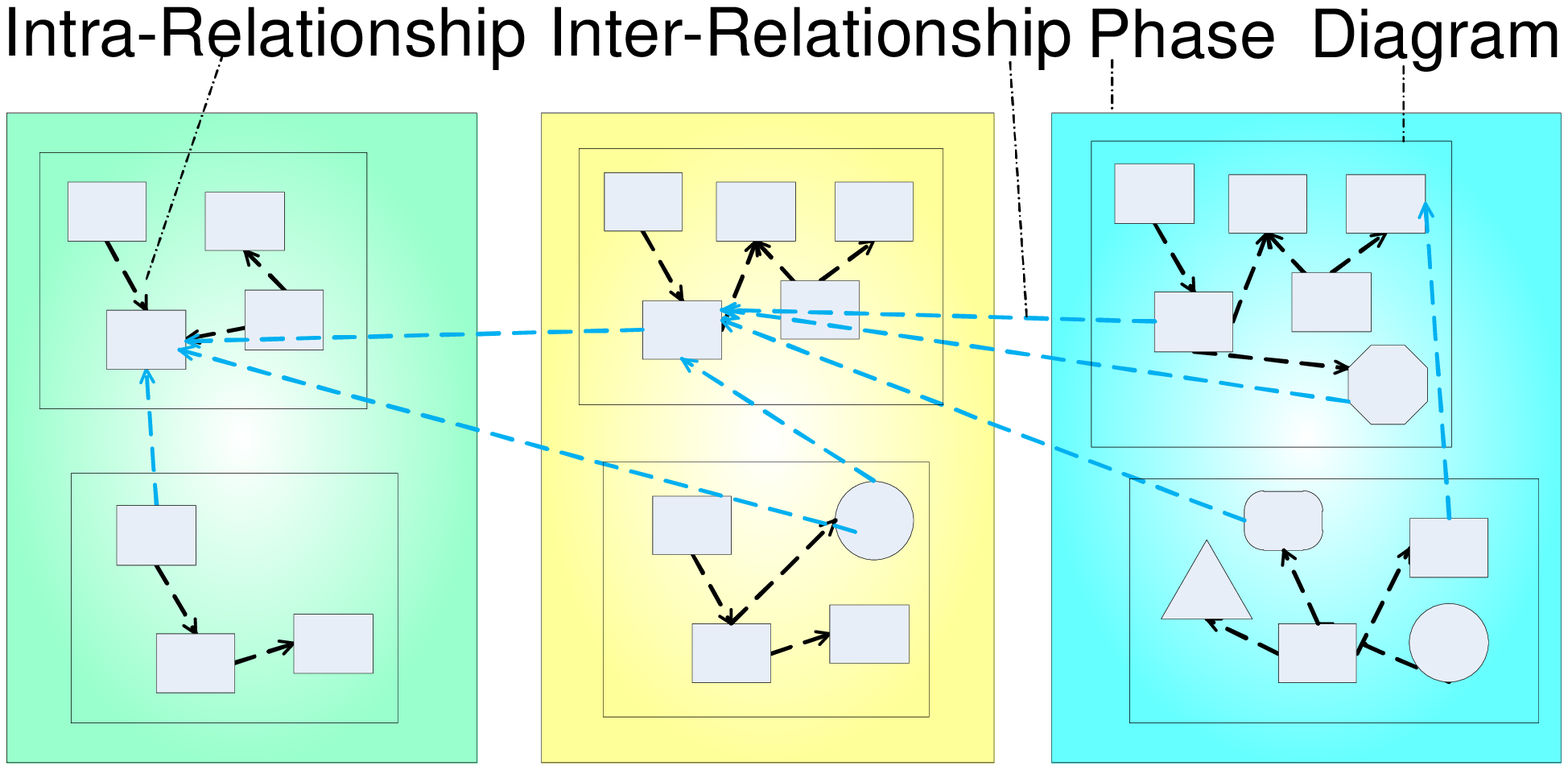}
\caption{Relationships among software artifacts }
\label{fig_relationship}
\end{figure}

% The appearance of an artifact is not limited to a diagram or a piece of code, but it may appear in many places during software development process under the form of a copy, an instance or even a new artifact but at different levels of abstraction. 
% 11th April When a worker create a UML diagram or a piece of source code, he also specifies explicitly relationships among artifacts inside this diagram or this program. However, there are other relationships besides explicit ones. Most artifacts will not appear only once in a diagram or a piece of code, but they may appear in many places during software development process in different forms such as a copy, an instance or even a new artifact but at different levels of abstraction, etc. 
%The relationships among different forms of an artifact are implicit but existed. 
% 11th April Based on these observations, we classify relationships among software artifacts into two groups (Fig. \ref{fig_relationship}). The first is \textit{intra-relationships}  that are relationships specified explicitly by workers when they specify a diagram or a program. Relationships among model elements in the same diagram or relationships among source code elements are intra-relationships. The second is \textit{inter-relationships}  that connect related artifacts in different diagrams or in different phases together.

We classify relationships among software artifacts into two groups (Fig. \ref{fig_relationship}). The first is \textit{intra-relationships}  that are relationships among model elements in the same diagram. The second is \textit{inter-relationships}  that connect related model elements in different diagrams or in different phases together.
% 11th April Based on this classification, we name dependency relationships in inter-relationship group as \textit{inter-dependency} and dependency relationships in intra-relationship group as \textit{intra-dependency} .  Because intra-dependency are specified explicitly by workers and stored in meta-data of software artifacts, there is a need for generating inter-dependency relationships automatically. Several studies have been made to generate dependencies among source code elements but ignore dependencies among model elements. Therefore, in this paper, we will pay attention to automatic generation of inter-dependencies among UML model elements, \textbf{Basic Dependency Relationships} (BDRs). Generating dependencies between UML model elements and source code will be left to future work.
Based on these classifications, we name dependency relationships in the inter-relationship group \textit{``inter-dependency"} and dependency relationships in intra-relationship group \textit{``intra-dependency"}. Examples of intra-dependency relationships are generalization, association (aggregation, composition), and usage dependency (call, instantiation, send, parameter). Examples of inter-dependency relationships are trace, refinement and derivation. In this paper, we will pay attention to automatic generation of inter-dependencies among UML model elements, named \textit{Basic Dependency Relationships} (BDRs). 

%First, we will build the Dependency Generator component. In this paper, we focus on generating Basic Dependency Relationships among UML model elements.
\begin{itemize}
\item 
%\textbf{Defining a Dependency Generation Model} that includes basic rules for identifying BDRs among UML model elements.
Define BDR types by analyzing dependency relationships defined in UML 1.5.
\item 
%\textbf{Developing a dependency generator} to generate the BDRs between UML model elements. The generator receives UML diagrams and process information, and returns the UML diagrams with BDRs attached.
% 11th April Developing a BDR Generation Engine to generate BDRs between UML model elements. This engine operates based on a set of rules for identifying BDRs among UML model elements. It will receive UML diagrams and process information, and returns the UML diagrams with BDRs attached. 
Develop a BDR Generation Engine to generate BDRs among UML model elements. This engine operates based on a set of rules for identifying BDRs among UML model elements. It receives UML model elements in UML diagrams and process information (phase names and phase orders), and returns the UML model elements with BDRs attached. 
\end{itemize}

%CSW Generator \& Management component is the most important component of our system, with two main functions: 

\subsubsection{CSW construction} Based on dependency relationships among software artifacts and change requests. By tracing dependency relationships, we can identify data elements (software artifact impacted by a change request) and the control flow (change order of impacted software artifacts) of a CSW. 
%Because change requests and dependencies are very complicated, automatically generated CSWs may not meet user's requirements. Therefore, our system allows workers to revise the generated CSWs to create correct workflows.
\subsubsection{Inconsistency Awareness} Consider the following points:
\begin{itemize}
\item 
Establish agreements on using CSWs among workers. Every worker will work based on a CSW. 
%Workers can supply extra information of their progress to support CSM in managing CSWs in the system. For example, workers can add some message describing their change purpose . They can also update states of software artifacts modified by activities in their CSWs such as start modifying, modifying, and finishing.
\item 
%Inform workers about the possibility of inconsistency occurrences, and support them in inconsistency resolution by collecting and analyzing information of CSWs in the system.
Inform workers about (potential) inconsistencies, and support them in inconsistency resolution by collecting and analyzing information of CSWs in the system.
\end{itemize}

\subsection{CSM Framework}

\begin{figure}[!t]
\centering
\includegraphics[scale = 0.45]{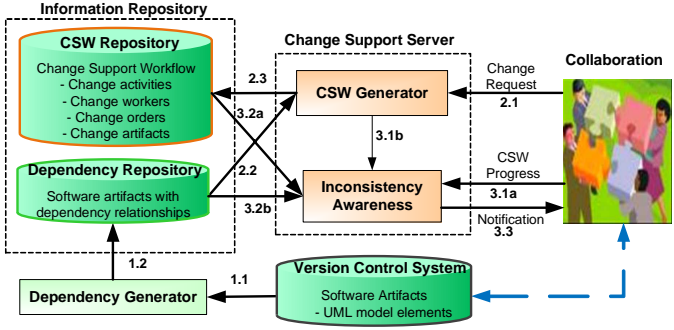}
\caption{Conceptual framework of CSM}
\label{fig_frameworkCSM}
\end{figure}

Fig. \ref{fig_frameworkCSM} describes the conceptual framework of the system with the following main components: 

\begin{itemize}
\item 
\textbf{Dependency Repository:} is a part of Information Repository. This repository contains UML model elements linked by dependency relationships.
\item 
\textbf{CSW Repository:} is a part of Information Repository. It contains information of CSWs executed in local workspaces of workers.
\item 
\textbf{Version Control System:} manages changes to software artifacts.
\item
\textbf{Dependency Generator:} generates dependency relationships among software artifacts in Version Control System and stores them in  Dependency Repository.
\item
\textbf{CSW Generator:} generates CSWs based on change requests and information in Dependency Repository, and stores them in CSW Repository.
\item
\textbf{Inconsistency Awareness:} notifies workers of the possibility of inconsistency based on collected information of CSWs at clients, and information in Dependency Repository and CSW Repository. 
\end{itemize}

The main processing flow of the system is as follows:
\begin{enumerate}
\item Generate dependency relationships by using Dependency Generator component (1.1, 1.2). 
\item Generate a CSW for each change request from workers by using CSW Generator component (2.1, 2.2, 2.3).
\item Notify the workers of the possibility of inconsistency in the newly generated CSWs and executing CSWs in the system using Inconsistency Awareness component (3.1a or 3.1b, 3.2a and 3.2b, 3.3).
\end{enumerate}
%\begin{figure}[!t]
%\centering
%\includegraphics[scale=0.45]{MainStepBuildingCSM.eps}
%\caption{Main steps in building main components of CSM  }
%\label{fig_mainStepsCSM}
%\end{figure}

%Fig. \ref{fig_mainStepsCSM} shows main tasks to build Dependency Generator component and CSW Generator \& Management component of CSM. 

\section{Dependency Generation}

%We classify relationships among software artifacts into two groups (Fig. \ref{fig_relationship}). The first is intra-relationships  that are relationships specified explicitly by workers when they specify a diagram or a program. Relationships among model elements in the same diagram or relationships among source code elements are intra-relationships. The second is inter-relationships  that connect artifacts in different diagrams or in different phases together. Basic Dependency Relationships (BDRs) belong to inter-relationship group . 

%In this section, we will define types of BDRs by analyzing dependency relationships of UML. We also present a method and a tool for automatic generation of BDRs from UML diagrams created during phases of a software development process \cite{ref_Kotani_journal}. 
In this section, we will define types of BDRs by analyzing dependency relationships of UML, and present a method and a tool for automatic generation of BDRs from UML diagrams created during the software development process \cite{ref_Kotani_journal}. 

\subsection{Basic Dependency Relationship}
UML 1.5 has defined `dependency' as a relationship between two elements in which a change to one element, the Target, may affect or supply information needed by the other element, the Source. Dependency has many varieties that represent different kinds of relationships: Abstraction, Binding, Permission, and Usage. In addition to these varieties, we realize that developers also create some implicit relationships among UML model elements: Copy, relationship among copies of an element but in different diagrams, and Inclusion, the whole-part relationship between an element and its internal members or between a diagram and its elements.  By analyzing the `dependency' of UML 1.5 and the dependencies generated implicitly by developers, we propose a set of new generable Basic Dependency Relationships (BDRs) between UML model elements: Exist Together, Information Sharing, Copy, and Concept (see Table \ref{tab_dependency_analysis}).

\subsubsection{Exist Together} The Source does not exist without the Target.
Exist Together is defined from Usage and Inclusion relationships. Usage and Inclusion relationships can be generated automatically by comparing the names of UML model elements or analyzing the whole-part relationship. Therefore, we can generate Exist Together relationships automatically if the names of model elements are comparable or the Inclusion relationship can be analyzed.

\subsubsection{Information Sharing}  Information of the Target is a part of information of the Source.
Information Sharing has been extracted from Binding, Permission, and Usage. Although Binding and Permission cannot be generated automatically, Usage can be generated automatically if the names of UML model elements are comparable. Therefore, when at least one name in the shared information group is comparable, we can generate Information Sharing relationship automatically.
\subsubsection{Copy}  Information of the Target and the Source is the same. 
This relationship is extracted from the Copy relationship generated implicitly by developers. Copy relationship can be generated automatically among UML diagram elements which are in the same phase, and have the same name and type.
\subsubsection{Concept} The Source and the Target represent the same concept but the Source is more concrete. 
Concept has been extracted from Abstraction. Abstraction can be generated automatically by comparing the names of UML model elements and using the process information. Therefore, we can generate Concept relationship automatically if process information is given and elements representing the same concept are  named similarly. 

\begin{table}[!t]
\centering
\caption{Dependency Analysis}
\scalebox{0.88}[0.9]{
\begin{tabular}{|c|c|c|} \hline
\textbf{UML 1.5 Dependency} & \textbf{BDR Types} & \textbf{Automatic Generation Method} \\ \hline
%\textbf{Dependency} &             &                  \\ \hline
Abstraction & Concept             & Process Information \\
				&			             & + Name Comparison \\ \hline
Binding     & Information Sharing & (None) \\ \hline
Permission	& Information Sharing & (None)\\ \hline
Usage	      & Exist Together      & Name comparison \\
            & Information Sharing & \\ \hline
Copy        & Copy                & Name comparison \\ \hline
Inclusion   & Exist Together      & Inclusion of UML description\\ \hline
\end{tabular}}
\label{tab_dependency_analysis}
\end{table}

\subsection{Dependency Generation Model}
\begin{figure}[!t]
\centering
\includegraphics[scale=0.45]{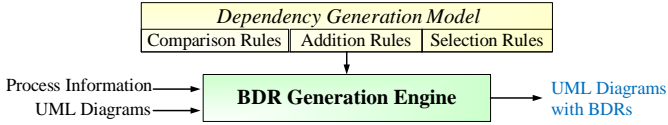}
\caption{Method for generating BDRs automatically}
\label{fig_BDRGenerationEngine}
\end{figure}
To generate the dependency relationships automatically, we define a Dependency Generation Model (DGM) consisting of comparison rules, addition rules, and selection rules. Comparison rules look for pairs of UML model elements that may have some BDRs, based on the similarity in names between UML model elements, and inclusion relationships between a diagram and its components. Addition rules identify BDR candidates that may be set to a pair of UML model elements. Regarding the information about phase, diagram, type and name of UML model elements, selection rules will choose one BDR from the BDR candidates found by addition rules to attach to the selected pair. Based on DGM, BDR Generation Engine accepts as input a group of UML diagrams and their process information (phase names and phase orders). Outputs will be these UML diagrams with newly added BDRs (see Fig. \ref{fig_BDRGenerationEngine}).

\subsubsection{Comparison rules}
Find pairs of UML elements to which BDRs may be attached. A BDR may exist between two UML model elements if they satisfy one of the following conditions:

\begin{itemize}
\item 
\textbf{Contained:} The name of the Target is included in the name of the Source, for example ``Elevator" and ``ElevatorControl".
\item
\textbf{Similar:} The name of the Target is similar to the name of the Source, for example ``FloorLampInterface" and ``FloorLampInterfaces".
\item 
\textbf{TypeSim:} The type of the Target is similar to the name of the Source, for example ``:FloorLampInterfaces" and ``FloorLampInterface".
\item 
\textbf{SimType:} The name of the Target is similar to the type of the Source, for example ``FloorLampInterface" and ``: FloorLampInterfaces".
\end{itemize}

Table \ref{tab_comparison_condition} shows suitable comparison conditions for each type of Target and Source.

\begin{table}[!t]
\centering
\caption{Comparison Conditions}
\scalebox{0.95}[0.95]{
\begin{tabular}{|c|c|c|} \hline
\textbf{Target}   & \textbf{Source}        & \textbf{Comparison Condition}  \\ \hline
%                  &                        & \textbf{ Condition} \\
Use case diagram	& Actor	                & Include \\ \hline
Use case diagram	& Use case               & Include \\ \hline
Use case	         & Class diagram	       & Contained \\ \hline
Use case	         & State chart diagram    & Contained \\ \hline
Use case	         & Collaboration diagram	 & Contained \\ \hline
Use case	         & Sequence diagram	    & Contained \\ \hline
Use case	         & Use case	             & Similar \\ \hline
Actor	            & Actor                  & Similar \\ \hline
Actor	            & Class                  & Similar \\ \hline
Actor	            & Object	                & SimType \\ \hline
Class diagram     & Class                  & Include \\ \hline
Class diagram     & Package                & Include \\ \hline
Package           & Class                  & Include \\ \hline
Package           & Package                & Similar \\ \hline
Class             & Package                & Similar \\ \hline
Class             & Object                 & SimType, Contained \\ \hline
Class             & State chart diagram    & Contained \\ \hline
Class             & Activity diagram       & Contained \\ \hline
Class             & Class                  & Similar, Include \\ \hline
Object diagram    & Object	                & Include \\ \hline
Object	         & Class	                & TypeSim \\ \hline
Object	         & State chart diagram	 & TypeSim \\ \hline
Object	         & Activity diagram	    & TypeSim \\ \hline
Object	         & Object	                & Similar, Include \\ \hline
Component diagram	& Component	             & Include \\ \hline
Component	      & Component	             & Similar \\ \hline
Deployment diagram   & Node	             & Include \\ \hline
Node	               & Node	             & Similar \\ \hline
State chart diagram	& State	             & Include \\ \hline
State	               & State	             & Similar, Include \\ \hline
Activity diagram	   & Action State 	    & Include \\ \hline
Action State	      & Action State	       & Similar, Include \\ \hline
Collaboration diagram& Object	             & Include \\ \hline
Sequence diagram	   & Object	             & Include \\ \hline
\end{tabular}}
\label{tab_comparison_condition}
\end{table}

\subsubsection{Addition rules} 
\begin{figure}[!t]
\centering
\includegraphics[scale = 0.45]{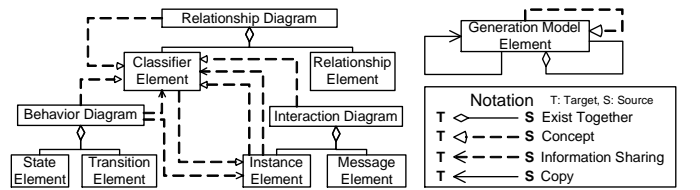}
\caption{Addition rules}
\label{fig_addition_rule}
\end{figure}

Identify BDRs which can exist between two UML model elements satisfying the comparison rules. We classify UML model elements into new categories which we call Generation Model Elements (See Table \ref{tab_GME}). We also define types of BDR which can be set between two Generation Model Elements (See Fig. \ref{fig_addition_rule}). Based on these instructions, we can find BDR candidates between any two UML model elements by mapping them to Generation Model Elements.

\begin{table}[!t]
\centering
\caption{GENERATION MODEL ELEMENTS}
\scalebox{0.99}[0.99]{
\begin{tabular}{|c|c|} \hline
\textbf{Generation Model Element} & \textbf{UML Model Element} \\ \hline
Classifier Element                & Actor, Use case, Class, Package, Node, \\
				                      & Component, Object (Object diagram) \\ \hline
Relationship Element              & Relation, Aggregation, Dependency,  \\ 
				                      & Generalization, Link \\ \hline
State Element                     & State, Action State\\ \hline
Transition Element	             & Transition, Event, Action \\ \hline
Instance Element                  & Object (Collaboration diagram, \\
                                  & Sequence diagram) \\ \hline
Message Element                   & Message \\ \hline
Relationship Diagram              & Use case diagram, Class diagram \\
				                      & Object diagram, Component diagram,  \\ 
                                  &     Deployment diagram, \\ \hline
Behavior Diagram	                & State chart diagram, Activity diagram \\ 
                                                                        \hline
Interaction Diagram	             & Sequence diagram, Collaboration diagram\\ 
                                                                        \hline
\end{tabular}}
\label{tab_GME}
\end{table}

\subsubsection{Selection rules}

\begin{table}[!t]
\setlength{\abovecaptionskip}{-2.5mm}
\caption{Selection rules}

\label{tab_addition_rule}

\begin{center}
\scalebox{0.77}[0.77]{
\begin{tabular}{|c|c|c|c|c|c|c|c|}
\hline
\multicolumn{2}{|c|}{}                                       & \multicolumn{4}{|c|}{Phase}                                                                                 &\multicolumn{2}{|c|}{}          \\ \cline{3-6}
\multicolumn{2}{|c|}{}                                       & \multicolumn{2}{|c|}{Same}                                        & Adjoining             & Separate        &\multicolumn{2}{|c|}{}          \\ \hline
\multirow{4}{35pt}{UML Element's Type} & \multirow{2}*{Same}      & \multirow{4}{25pt}{Exist Together}&Copy                                & \multirow{4}*{Concept}& \multirow{4}*{-}&Same      & \multirow{4}*{Name} \\ \cline{4-4}\cline{7-7}
                                  &                          &                              & -                                  &                       &                 &Different &                     \\ \cline{2-2}\cline{4-4}\cline{7-7}
                                  & \multirow{2}*{Different} &                              & \multirow{2}{35pt}{Information Sharing} &                       &                 &Same      &                     \\ \cline{7-7}
                                  &                          &                              &                                    &                       &                 &Different &                     \\ \hline
\multicolumn{2}{|c|}{}                                       &Same                          &\multicolumn{3}{|c|}{Different}                                               &\multicolumn{2}{|c|}{}          \\ \cline{3-6}
\multicolumn{2}{|c|}{}                                       &\multicolumn{4}{|c|}{Diagram}                                                                                &\multicolumn{2}{|c|}{}          \\ \hline
\end{tabular}
}
\end{center}
\end{table}

%\begin{table}[!t]
%\centering
%\caption{Selection rules}
%\includegraphics[scale = 0.95]{SelectionRule.eps}
%\label{tab_addition_rule}
%\end{table}

Decide on the BDR candidates obtained after applying the addition rules to two UML model elements satisfying the comparison rules. In Table \ref{tab_addition_rule}, we describe how to choose BDR by using information of names, types, diagram and phases of UML elements.
\begin{itemize}
\item 
\textbf{Information Sharing:} Two UML elements are in the same phase but in different diagrams, and are of different types.
\item
\textbf{Copy:} Two UML elements are in the same phase but in different diagrams. They have the same name and are the same type.
\item 
\textbf{Concept:} Two UML elements are in adjoining phases.
\item
\textbf{Exist Together:} Two UML elements are in the same diagram, and certainly in the same phase.
\end{itemize}

\subsection{Automatic BDR Generation}

BDR generation includes five steps:
\begin{enumerate}
\item 
\textbf{Extraction of potential UML elements:} Search pairs of UML model elements satisfying conditions of the comparison rules.
\item
\textbf{Retrieval of Generation Model Elements:} Find Generation Model Elements corresponding to these UML elements.
\item 
\textbf{BDR Candidate Identification:} List all BDR candidates between pairs of UML elements based on addition rules.
\item
\textbf{BDR Selection:} Choose a suitable BDR for each pair of UML elements based on selection rules.
\item
\textbf{BDR Addition:} Add the selected BDRs to the corresponding pairs.
\end{enumerate}

\begin{figure}[!t]
\centering
\includegraphics[scale = 0.42]{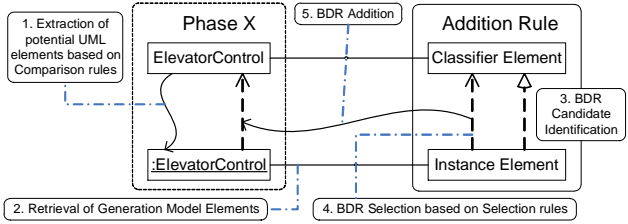}
\caption{Example of BDR generation}
\label{fig_example_BDRgeneration}
\end{figure}

Fig. \ref{fig_example_BDRgeneration} shows an example of automatic BDRs generation. Assume that we have some UML artifacts including the class ``ElevatorControl" and the object ``:ElevatorControl" which are in the same phase. Based on the comparison rules, these two UML elements satisfy the SimType condition. Therefore, we perform the second step, mapping these artifacts to Generation Model Elements. Using Table \ref{tab_GME}, Generation Model Elements of the class ``ElevatorControl" and the object ``:ElevatorControl" are Classifier element and Instance element, respectively. Next, the addition rules are applied. The diagram in Fig. \ref{fig_addition_rule} shows that there are two candidates for the BDR between the class ``ElevatorControl" and the object ``:ElevatorControl": Information Sharing and Concept. Because both artifacts are in the same phase, we decide on Information Sharing for the BDR between these artifacts. Finally, information about a BDR, Information Sharing, with the class ``ElevatorControl" at the Target and the object ``:ElevatorControl" at the Source is added to the system.

%\section{Prototype}
\subsection{Impact Analysis Tool}

\begin{figure}[!t]
\centering
\includegraphics[scale = 0.4]{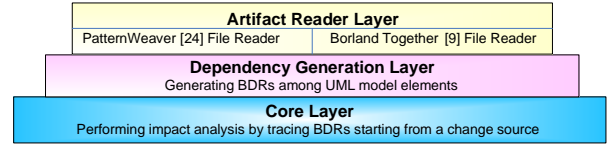}
\caption{Framework of the Impact Analysis Tool}
\label{fig_framework_impact_analysis_tool}
\end{figure}

\begin{figure}[!t]
\centering
\includegraphics[scale = 0.45]{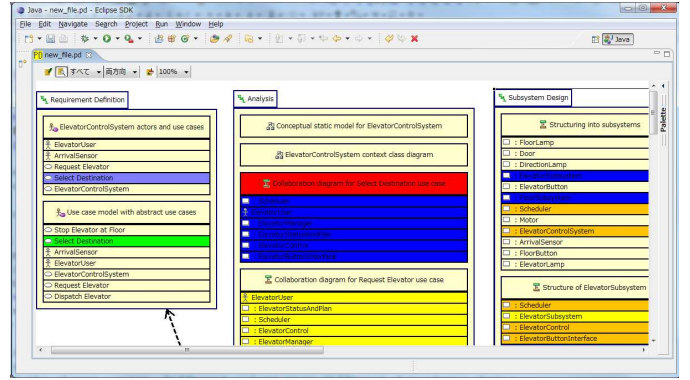}
\caption{Main window of the Impact Analysis Tool}
\label{fig_window}
\end{figure}

%We have developed a first prototype of CSS that implements dependency generator component and helps to generate a dependency graph starting from a change source. The prototype is developed as a plugin of Eclipse with three-layer framework (See Fig. \ref{fig_framework_CSS_Prototype}). 

We have developed an impact analysis tool that implements dependency generator component and performs impact analysis process starting from a change root.
%based on generated BDRs. 
This tool is developed as a plugin of Eclipse with three-layer framework (See Fig. \ref{fig_framework_impact_analysis_tool}). We have performed two case studies to evaluate our method. The precision of the generated BDRs is from 92.3\% to 93.3\%, and the recall is from 83.7\% to 87.7\% \cite{ref_Kotani_journal}.

%Fig. \ref{fig_window} shows a screen shot of the CSS prototype when we use impact analysis function to generate dependency graph of a change request by choosing the change source. Different colors mean different dependency chains.

%11th April Fig. \ref{fig_window} shows a screen shot of the impact analysis tool when we use impact analysis function to generate dependency graph of a change request by choosing the change source. Different colors mean different dependency chains.
Fig. \ref{fig_window} shows a screen shot of the tool when we find impact elements of a change request by choosing the change root. Different colors mean different dependency chains.

\section{Change Support Workflow Generation}
%In this section, we will present in detail our method for solving issues in building CSW Generation \& Management Component including CSW generation, inconsistency detection and resolution.
In this section, we will present the way to generate CSWs based on dependency relationships among artifacts.
\subsection{CSW Definition}
As a workflow, a CSW must contain basic information such as change activities, software artifacts accessed by change activities, and change orders. 
In addition, to support inconsistency detection, we need to store active intervals of activities in a CSW. Regarding access control, information of the change worker associated with an activity will be recorded. 
Below is a formal definition of CSW.

\underline{\textbf{Definition 1:}} A Change Support Workflow is a tuple $w =$  $<id, A, F, D, T, W, GD, GW>$ where:
\begin{itemize}

\item $id$ is the workflow identifier.
\item $A$ is a set of activities.
\item $F \subseteq (A \times A) $ is a set of arcs (flow relation) that represent the orders of change activities.
\item $D$ is a set of software artifacts accessed by activities of CSW.
\item $T = \{r, w\}$ is a set of tasks on software artifacts (r: \textit{\textbf{r}ead}, w: \textit{\textbf{w}rite}). \textit{\textbf{r}ead} means that this artifact is used to implement a change on another artifact. \textit{\textbf{w}rite} means that this artifact will be changed. 
\item $W$ is a set of workers who execute activities of CSW.
\item $GD: A \times T\rightarrow 2^D$ is a function that returns a set of artifacts associated with an activity and a task.
\item $GW: A\rightarrow 2^D$ is a function that returns the workers associated with an activity of CSW.
\item $GT: A \rightarrow ((R^+ \cup \infty)  \times (R^+ \cup \infty))$ is a time interval function that returns the start time and the finish time of  an activity. $R^+$ is the set of all positive real numbers. $\infty$ denotes an undecided start time or finish time. 
\end{itemize}
%\end{enumerate} 

%\subsection{Dependency Graph}
\subsection{CSW Generation}
When a worker makes a change to an artifact, \textit{change root}, this change may affect artifacts that have BDRs with the change root, and require extra modifications. By following BDRs among artifacts, we can find all potentially impacted elements. 

Here is a general algorithm for identifying \textit{dependency graph} starting from a \textit{root artifact}. Dependency graph is a directed graph where vertexes are potentially impacted artifacts, and edges are BDRs among the potentially impacted artifacts. An edge $e = (x; y)$ is considered to be directed from node x to node y if there is a BDR with the Target y and the Source x. %Source vertex is the source artifact itself.
\begin{enumerate}
\item The initial set of vertexes is the root artifact itself.
\item For each artifact in the vertex set, if there is a BDR having this artifact as the Target, add the Source element to the set if it is not already in the set, and add an edge directed from the Source to the Target.
\item Repeat Step 2 until all vertexes are examined and no new artifact appears.
\end{enumerate}

\begin{figure}[!t]
\centering
\includegraphics[scale = 0.45]{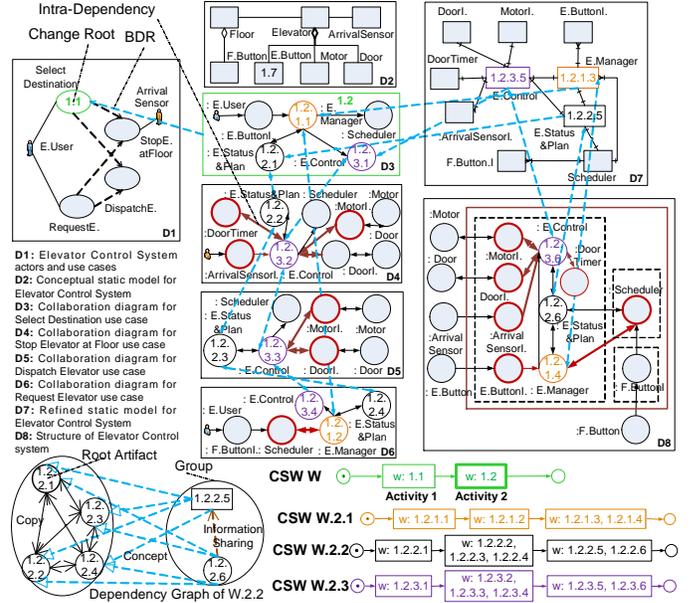}
\caption{Example of generating CSWs}
\label{fig_example_generating_CSW}
\end{figure}

The most straightforward way to generate a CSW from a dependency graph is to use the same structure as in the dependency graph. Each artifact is mapped to a new activity, and each BDR between two artifacts becomes an arc connecting two corresponding activities. However this method is just suitable for a very sparse dependency graph whereas a dependency graph is a dense graph in reality (See an example of a dependency graph at the bottom left corner of Fig. \ref{fig_example_generating_CSW}). Therefore, the structure of a CSW generated by this method is not a good formulation for change support, because several artifacts should be examined together. So we use a grouping technique to identify groups of strongly related artifacts and map each group to an activity in a CSW. A group contains artifacts connected together by Copy or Information Sharing relationships.
%12th April Following is the algorithm to generate a CSW by assigning artifacts in the dependency graph to activities in the CSW based on BDR types. The generated CSW includes information about elements  A, F, D, T, and GD of a CSW. Elements related to workers, W and GW, are decided by the some designers or project managers. Regarding the time interval, GT, we will collect this information during the workflow execution.
\begin{itemize}
%12th April \item Assign the head artifact to the \textit{\textbf{w}rite} data set of the first activity in the CSW. 
\item Put artifacts connected by Copy or Information Sharing relationships into a group.
\item Find the group containing the root artifact. Assign the root artifact to the \textit{\textbf{w}rite} data set of the first activity in the CSW, and the remaining artifacts to the \textit{\textbf{w}rite} data set of the second activity of the CSW.
 
%12th April\item If there is a Copy or an Information Sharing between an unassigned artifact and an assigned artifact, put this unassigned artifact to the  \textit{\textbf{w}rite} data set of the same activity with the assigned artifact. However, if the assigned artifact is the head artifact, put the unassigned artifact to the second activity of the CSW instead of the first activity.

%12th April\item If there is a Concept between an unassigned artifact and an assigned artifact, put this unassigned artifact to the  \textit{\textbf{w}rite} data set of the activity following the activity of the assigned artifact.
\item For the remaining groups, create the same number of new activities so that each new activity will receive each corresponding group as its \textit{\textbf{w}rite} data set. 
\item If there is at least one Concept relationship between two artifacts in two different activities, create an arc from the activity containing the Target to the activity containing the Source. 

%12th April\item If \textit{\textbf{w}rite} data set of an activity contains an artifact, a diagram that is the Target element of at least one Exist Together, this activity will be classified as a \textit{composite activity}. A composite activity can include many CSWs whose head artifacts are the Target elements of these Exist Together relationships.
\item If the \textit{\textbf{w}rite} data set of an activity contains an artifact, i.e. a diagram, that is the Target element of at least one Exist Together, this activity will be classified as a \textit{composite activity}. A composite activity can include many CSWs whose root artifacts are the Source elements of these Exist Together relationships.

\end{itemize}

Based on Definition 1, the generated CSW will include information about elements  A, F, D, T, and GD of a CSW. Elements related to workers, W and GW, are decided by designers or project managers. The time interval, GT, will be collected during workflow execution.

Fig. \ref{fig_example_generating_CSW} describes an excerpt of a non-distributed Elevator Control System [22] with some important diagrams in the requirement definition phase (D1), analysis phase (D2, D3, D4 and D5) and subsystem design phase (D7 and D8). We assume that there is a change request on Select Destination use case, Artifact 1.1. For simplicity, we just show important BDRs among artifacts. Applying the above algorithm with the change root as Artifact 1.1, we have  CSW W with two activities in which Activity 2 will modify Artifact 1.2, diagram D3. We assume that this change may affect three artifacts inside this diagram, 1.2.1.1, 1.2.2.1, and 1.2.3.1. Therefore, from the composite activity 2, three CSWs are built with these root artifacts: 1.2.1.1, 1.2.2.1, and 1.2.3.1 respectively. These three CSWs are also considered as three branches originating from the composite Activity 2. 
%All these four CSWs are classified as $1^{st}$ Grade CSWs. Because of intra-dependency relationships, changes caused by these  $1^{st}$ Grade CSWs can spread to other artifacts marked in the red color. $2^{nd}$ Grade CSWs can be built based on these artifacts and artifacts connected to these artifacts by BDRs.

\subsection{CSW Execution Control}

%12th April A CSW can be in one of three states: planning, executing and finished. A \textit{planning workflow} is a CSW at build time that is generated automatically by the CSW Generator component. In a planning workflow, some contents may be missing and some change orders are not decided yet. Different from planning workflow, an \textit{executing workflow} is a CSW that represents what really happens at runtime by a change worker based on a planning workflow. When an executing workflow finishes its last activity, it becomes a \textit{finished workflow}.

A CSW can be in one of three states: planning, executing, and finished. A CSW is in the planning state when it is generated automatically by the CSW Generator component at build time (CSW Generation Section). When a change worker starts the \textit{planning workflow}, this workflow will move to the executing state. When the change worker finishes the last activity of the \textit{executing workflow}, the CSW will be in the finished state, a \textit{finished workflow}.

When a CSW moves from the planning state to the executing state, some values of the workflow may be modified by workers, such as worker assignment. Also, the change orders of artifacts which are assigned to the same activity in the planning phase, are decided. Start time and finish time of each activity can also be  identified. 

However, the most important thing that happens at run time is the spread of dynamic change. Artifacts in these CSWs are not isolated, but are connected with  artifacts in the same diagram by intra-dependency relationships. Because of these relationships, when these CSWs are executed, changes on artifacts being modified by activities in these CSWs will spread to the intra-dependency related artifacts. In the example described in Fig. \ref{fig_example_generating_CSW}, the artifacts which may be affected when CSWs W.2.1, W.2.2, and W.3.3 start executing  are marked with thick red borders. Therefore, we need to dynamically generate new CSWs to meet the arising changes by using the same process described in the previous section. We call the original CSWs \textit{main CSWs}, and the newly arising CSWs \textit{sub-CSWs}.

\begin{figure}[!t]
\centering
\includegraphics[scale = 0.45]{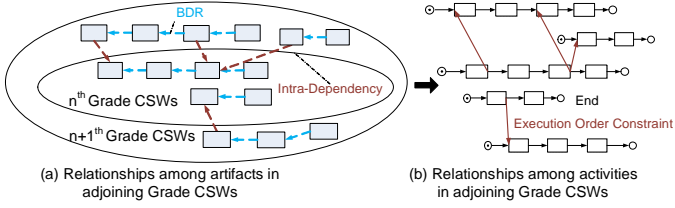}
\caption{Relationships among CSWs in adjoining grades}
\label{fig_relationships_among_adjoining_CSWs}
\end{figure}

To manage CSWs of a change request easily, we classify these CSWs into different grades, starting from $1^{st}$ grade. $1^{st}$ Grade CSWs are the main CSWs. CSWs from $2^{nd}$ Grade and higher are sub-CSWs. $2^{nd}$ Grade CSWs are built based on $1^{st}$ Grade CSWs. A $2^{nd}$ Grade CSW is a CSW of which the root artifact has intra-dependencies with artifacts in the $1^{st}$ Grade CSW, and has not yet appeared in any existing CSWs. After identifying the root artifact, we can find the elements of a new CSW by tracing BDRs starting from this root artifact. In general, $n+1^{th}$ Grade CSWs will be built based on $n^{th}$ Grade CSWs in the same way (Fig. \ref{fig_relationships_among_adjoining_CSWs}). 
% 13th April Therefore, we can generate all CSWs of a change request following the bottom-up approach.

%\begin{figure}[!t]
%\centering
%\includegraphics[scale = 0.35]{GeneratingCSWApproach.eps}
%\caption{Approach to generate CSWs}
%\label{fig_generatingBDR_approach}
%\end{figure}

%\subsection{Scheduling CSWs of a change request}

%\begin{figure}[!t]
%\centering
%\includegraphics[scale = 0.4]{Pipeline.eps}
%\caption{Main CSW and its sub CSWs}
%\label{fig_pipeline}
%\end{figure}

%11th April For a change request, we may have many CSWs in different grades. To ensure that changes are executed in an exact and effective manner, we schedule these CSWs in a pipeline mode (Fig. \ref{fig_pipeline}). If two activities in two adjoining grade CSWs have an intra-dependency relationship between their artifacts, the activity in the higher grade CSW should be executed after the activity in the lower grade CSW. Because all these CSWs are constructed to fulfill a change request, small change executed by each CSW should be consistent with the original change request. Worker of a CSW should implement changes with respect to the original change request and CSWs in lower grades, especially CSWs to which his CSW is directly connected by intra-dependency relationships. A change in a higher level CSW should not cause a change to artifacts already belonging to lower level CSWs to avoid a loop of CSWs. CSM will support change workers who execute CSWs of the same change request by notifying them joint points and force them to follow the constraints, intra-dependency relationships, between their own CSWs and others.

To ensure that changes are executed in an exact and effective manner, we schedule CSWs of a change request in a pipeline mode. If two artifacts in two adjoining grade CSWs have an intra-dependency relationship, the activity in the higher grade CSW should be executed after the activity in the lower grade CSW (See Fig. \ref{fig_relationships_among_adjoining_CSWs}b). 
%13th April Because all these CSWs are constructed to fulfill a change request, small changes executed by each CSW should be consistent with the original change request. A worker of a CSW should implement changes with respect to the original change request and CSWs in lower grades, especially CSWs to which his CSW is directly connected by intra-dependency relationships. A change in a higher level CSW should not cause a change to artifacts already belonging to lower level CSWs, to avoid a loop of CSWs. CSM will support change workers who execute CSWs of the same change request by notifying them joint points and force them to follow the execution order constraints, the intra-dependency relationships, between their own CSWs and others.
CSM supports the change workers who execute CSWs of the same change request, by notifying them of common points, and forcing them to follow the execution order constraints, the intra-dependency relationships, between their own CSWs and others.

%11th April Due to intra-dependencies and BDRs, most artifacts are connected together. This means that a change can spread to most artifacts of the system starting from a change source. Change order is another problem. Intuitively, dependency relationships among impacted elements represent the order of change. Nevertheless, there may be more than one path connected by two artifacts. In addition, dependencies are not necessary transitive; it depends on artifact contents and change requirements. Because we cannot cover all these problems, change workers must give final decisions based on CSWs generated by our model such as identifying the stop point of a CSW, change orders, etc. 

\section{Inconsistency Awareness}
%In a distributed collaborative environment, there may be many CSWs executed at the same time and they may share some software artifacts. Therefore, CSM will consider not only the CSW construction but also the inconsistency of changes on shared artifacts or dependent artifacts by different CSWs.

%\begin{figure}[!b]
%\centering
%\includegraphics[scale = 0.44]{Conflict.eps}
%\caption{Conflicts among CSWs}
%\label{fig_conflict}
%\end{figure}
 	
Most previous works in inconsistency awareness concentrated on code conflict, a kind of inconsistency caused by concurrent changes on shared artifacts.
%(See Fig. \ref{fig_conflict}). 
In \cite{ref_Indirect_Conflict_Awareness_Sarma}, the authors classified conflicts into two types: direct conflict and indirect conflict. Direct conflicts are caused by concurrent changes to the same artifact.  Indirect conflicts are caused by changes in one artifact affecting concurrent changes in another artifact. In general, indirect conflicts are more dangerous because they are detected late. CSM will consider both types of conflicts. We also pay attention to other types of inconsistencies occurring when activities in concurrent CSWs modify artifacts connected by some dependency relationships.

Like most other systems, CSM uses Version Control Systems (VCSs) to manage artifacts, and to support collaboration and parallel work. VCSs can detect direct conflicts by comparing check-in versions. However, VCSs cannot help in the case of indirect conflicts, because changes are implemented on different artifacts. Even if in the case of direct conflicts, VCSs detect them at check-in time when all changes have finished. This is a waste of time and effort. These inconsistencies should be detected as soon as possible. Our method is to detect (potential) inconsistencies at both build time and runtime. 

\begin{figure}[!t]
\centering
\includegraphics[scale = 0.3]{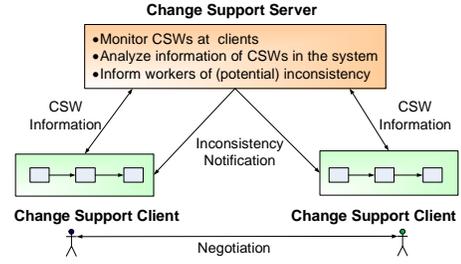}
\caption{Approach to handling inconsistencies }
\label{fig_handling_inconsistency_approach}
\end{figure}

Fig. \ref{fig_handling_inconsistency_approach} shows our approach to handling inconsistency using client-server architecture. Change workers use Change Support Clients to interact with their CSWs. A Change Support Server monitors CSWs at clients through Change Support Clients, analyzes collected information, and notifies change workers of (potential) inconsistencies. 
%Change workers will need some negotiations to find a solution for the inconsistencies.
Change workers will need to negotiate to find resolutions for the inconsistencies.
\subsection{Inconsistency Detection}

Potential inconsistencies at build time are detected by checking the existence of shared artifacts among planning workflows, or among a planning workflow and executing workflows. Right after generating a new planning workflow, the system will check potential inconsistencies between the new workflow and other planning workflows or executing workflows. If there are shared artifacts between two planning workflows, there is a chance for the related workers to cooperate to reconsider their change requests and CSWs. If shared artifacts are detected between the new workflow and an executing workflow, new workflow should be delayed until the executing workflow has finished. Another candidate method is to use negotiation, as in the previous case.

\begin{figure}[!t]
\centering
\includegraphics[scale = 0.45]{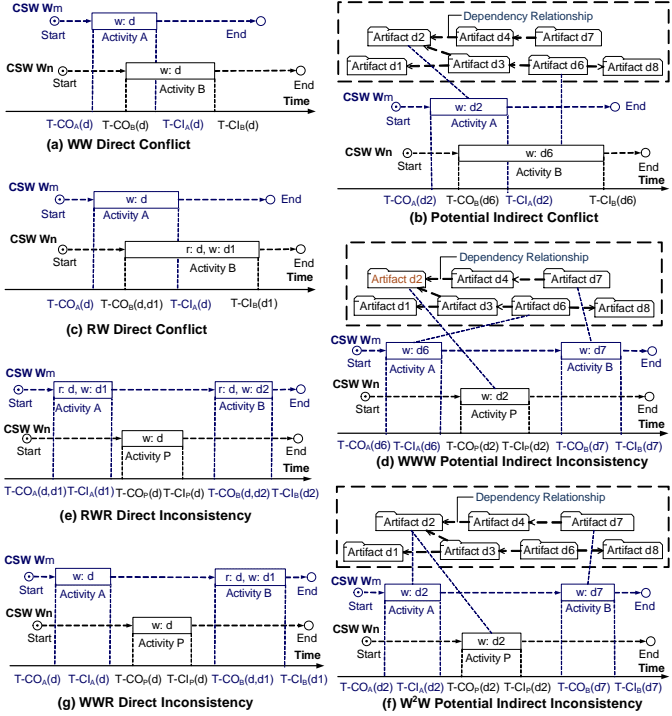}
\caption{Inconsistency patterns }
\label{fig_inconsistency_pattern}
\end{figure}

To detect inconsistencies at runtime, we identify patterns of (potential) inconsistencies among executing workflows. These patterns are special cases of Unintentional Change in In-use Data (UCID) patterns that we have presented in \cite{ref_UCID_PNSE}, \cite{ref_UCID_SERP}. We assume that a worker will check out the latest versions of all necessary artifacts at the beginning of an activity, and check in modified artifacts at the end of each activity for all activities in his CSW. 

%Fig. 12 shows our approach in handling inconsistency. Change Support Server will collect information of local CSWs, analyze them and notify change workers of inconsistencies or potential inconsistencies. 

The following notations are used in the definitions of inconsistency patterns:

\begin{itemize}
\item $CO_A(d)$: Activity A checks out the latest version of artifact d.
\item $T-CO_A(d)$: Point in time when activity A checks out artifact d.
\item $V-CO_A(d)$: Version of d when it is checked out by activity A.
\item $CI_A(d)$: Activity A checks in artifact d.
\item $T-CI_A(d)$: Point in time when activity A checks in artifact d.
\item $V-CI_A(d)$: Version of d when it is checked in by activity A.
\item $[T-CO_A, T-CI_A]$: Active Interval of activity A.
%\item Two activities in the same workflow means the ids of their CSWs are the same or their CSWs are designed for the same change request. 
%\item Two activities in different workflows means their CSWs are designed for different change requests 
\end{itemize}

\textbf{\underline{Pattern 1}: WW Direct Conflict} occurs when two activities in different CSWs \textit{concurrently} modify (\textit{\textbf{w}rite}) \textit{the same version of an artifact}, and create two new conflicting versions.

Fig. \ref{fig_inconsistency_pattern}a describes a WW Direct Conflict between two activities A and B in different CSWs:
\begin{itemize}
\item A and B have overlapping Active Intervals: $[T-CO_A(d), T-CI_A(d)]$ $\cap  [T-CO_B(d), T-CI_B(d)] \ne \emptyset$
\item Versions of the shared artifact between A and B at check-out time are the same: $V-CO_A(d) = V-CO_B(d)$
\item Versions of the shared artifact between A and B at check-in time are different: $V-CI_A(d) \ne V-CI_B(d)$
\end{itemize}

Based on Fig. \ref{fig_inconsistency_pattern}a, we can detect the possibility of a WW Direct Conflict at the start time of activity B, instead of at the finish time of B, when change worker checks in the shared artifact.

\textbf{\underline{Pattern 2}: Potential Indirect Conflict} occurs when two activities in different CSWs \textit{concurrently} modify \textit{two different artifacts} that are \textit{connected by dependency relationships} (BDRs or intra-dependency relationships). 

Fig. \ref{fig_inconsistency_pattern}b describes a Potential Indirect Conflict between two concurrent activities A and B in different CSWs:
\begin{itemize}
\item A and B have overlapping Active Intervals: $[T-CO_A(d_2),T-CI_A(d_2)] \cap  [T-CO_B(d_6), T-CI_B(d_6)] \ne \emptyset$
\item $d_2$ can be reached from $d_6$ by dependency relationships
\end{itemize}

Based on Fig. \ref{fig_inconsistency_pattern}b, we can detect this risk at the start time of activity B when change workers checks out the artifact $d_6$ that depends on the artifact  $d_2$ being modified by the concurrent activity A.

We call this pattern potential indirect conflict because conflict just happens between two artifacts if modification of an artifact is based on the content of the other artifact. This special situation is called \textbf{RW Direct Conflict}.

\textbf{RW Direct Conflict} occurs when an activity A uses (\textit{\textbf{r}ead})  a version of an artifact d to modify (\textit{\textbf{w}rite}) another artifact, and an activity B in a different CSW concurrently modifies (\textit{\textbf{w}rites})  the same version of d with A.

Fig. \ref{fig_inconsistency_pattern}c describes an RW Direct Conflict between two activities A and B in different CSWs. B will modify (\textit{\textbf{w}rite}) $d_1$ based on $d$.
\begin{itemize}
\item A and B have overlapping Active Intervals: $[T-CO_A(d), T-CI_A(d)] \cap  [T-CO_B(d,d_1), T-CI_B(d_1)] \ne \emptyset$
\item Versions of the shared artifact between A and B at check-out time are the same: $V-CO_A(d) = V-CO_B(d)$
\item The version of the shared artifact at check-in time of A is different from the version at check-out time of B: $V-CI_A(d) \ne  V-CO_B(d)$
\end{itemize}

\textbf{\underline{Pattern 3:} WWW Potential Indirect Inconsistency} occurs between two artifacts that are connected to the same artifact by dependency relationships (BDRs or intra-dependency relationships) and are modified (\textit{\textbf{w}ritten}) by two activities in the same CSW, because an activity in a different CSW (\textit{\textbf{w}rote}) to this shared artifact sometimes during the interval between these two activities.

Fig. \ref{fig_inconsistency_pattern}d describes a WWW Potential Indirect Inconsistency between three activities A, B, and P in which A and B are in the same CSW, and P is in a different CSW:
\begin{itemize}
\item P happens sometimes during the interval between A and B: $[T-CO_P(d_2), T-CI_P(d_2)] \subset [T-CI_A(d_6), T-CO_B(d_7)]$
\item $d_2$ can be reached from $d_6$ and $d_7$ by dependency relationships
\end{itemize}

Based on Fig. \ref{fig_inconsistency_pattern}d, we can detect this potential inconsistency at the start time of activity B or activity P.

We name this pattern potential indirect inconsistency because inconsistency just happens between two artifacts in the same workflow if modifications of both artifacts are based on the content of the same artifact. This special situation is also called \textbf{RWR Direct Inconsistency}.

\textbf{RWR Direct Inconsistency} occurs when two activities in the same CSW use (\textit{\textbf{r}ead}) different versions of an artifact to modify (\textit{\textbf{w}rite}) other artifacts, which are different from the initial plan of their CSW, because an activity in a different CSW (\textit{\textbf{w}rote}) to this shared artifact sometimes during the interval between these two activities.

Fig. \ref{fig_inconsistency_pattern}e describes a \textbf{RWR Direct Inconsistency} between three activities A, B and P in which A and B are in the same CSW, and P is in a different CSW. A and B will modify (\textit{\textbf{w}rite}) $d_1$ and $d_2$,  respectively based on d.
\begin{itemize}
\item P happens sometimes during the interval  between A and B: $[T-CO_P(d), T-CI_P(d)] \subset  [T-CI_A(d_1), T-CO_B(d,d_2)]$ 
\item Version of the shared artifact at check-in time of A is the input of P: $V-CI_A(d) = V-CO_P(d)$
\item Version of the shared artifact at check-in time of P is the input of B: $V-CI_P(d) = V-CO_B(d)$
\end{itemize}

\textbf{\underline{Pattern 4:} $W^2W$ Potential Indirect Inconsistency} occurs when two activities in the same CSW modify (\textit{\textbf{w}rite}) two artifacts, of which the artifact modified by the later activity is connected to the previous artifact by dependency relationships (BDRs or intra-dependency relationships), and another activity in a different CSW (\textit{\textbf{w}rote}) to the previous artifact sometimes during the interval between these two activities.

Fig. \ref{fig_inconsistency_pattern}f describes a $W^2W$ Potential Indirect Inconsistency between three activities A, B, and P in which A and B are in the same CSW, and P is in a different CSW:
\begin{itemize}
\item P happens sometimes during the interval between A and B: $[T-CO_P(d_2), T-CI_P(d_2)] \subset  [T-CI_A(d_2), T-CO_B(d_7)]$
\item $d_2$ can be reached from $d_7$ by dependency relationships
\end{itemize}

Based on Fig. \ref{fig_inconsistency_pattern}f, we can detect this type of potential inconsistency at the start time of activity B or activity P.

We name this pattern potential indirect inconsistency because inconsistency just happens between two artifacts in the same workflow if modifications of both artifacts are based on the content of the same artifact. This special situation is also called \textbf{WWR Direct Inconsistency}.

\textbf{WWR Direct Inconsistency} occurs when an activity of a CSW uses (\textit{\textbf{r}eads}) a different version of an artifact, instead of the version created by another activity in the same CSW, to modify (\textit{\textbf{w}rite}) another artifact, because an activity in a different CSW (\textit{\textbf{w}rote}) to this shared artifact sometimes during the interval between these two activities.

Fig. \ref{fig_inconsistency_pattern}g describes a WWR Direct Inconsistency between three activities A, B, and P in which A and B are in the same CSW, and P is in a different CSW. B will modify (\textit{\textbf{w}rite}) $d_1$ based on d.
\begin{itemize}
\item P happens sometimes during the interval between A and B: $[T-CO_P(d), T-CI_P(d)] \subset  [T-CI_A(d), T-CO_B(d, d_1)]$
\item Version of the shared artifact at check-in time of A is the input of P: $V-CI_A(d) = V-CO_P(d)$
\item Version of the shared artifact at check-in time of P is the input of B:  $V-CI_P(d) = V-CO_B(d)$
\end{itemize}
\subsection{Inconsistency Resolution}

Resolving (potential) inconsistencies among CSWs is not simple because different workers design different CSWs for different change requests, and a designer may know nothing about the work of others. Therefore, the cooperation of change workers is the most important factor. When receiving a warning of inconsistencies or potential inconsistencies from CSM, the related workers will contact with each other to conduct a negotiation. Face to face discussion, email, phone, instant messenger, etc. can be their communication means. Below are some methods for fixing inconsistencies which change workers can consider in their negotiation:
\begin{itemize}
\item Use a fine-grain work approach. CSWs can still work concurrently if they modify different parts of inconsistency-related artifacts.
\item Create a new change request that is a combination of change requests implemented by inconsistency-related CSWs. We will replace these inconsistency-related CSWs with the new CSWs that implement the new change request. This method can apply to potential inconsistencies between planning workflows.
\item Merge inconsistency-related parts of CSWs to create a new workflow.
\end{itemize}

\section{Related Work}

In the change support field, much previous work focused on change impact analysis on source code \cite{ref_change_impact_analysis}. By generating and managing CSWs, our CSM aims to support not only impact analysis, but also change planning and change execution.

Regarding collaborative inconsistency, most previous studies are about code conflicts caused by concurrent changes of different developers. Traditional approach uses a version control system such as CVS \cite{ref_CVS} or Subversion \cite{ref_Subversion} in conjunction with the programming environment to address the problem of concurrent accesses. An issue with this approach is that conflicts are detected at check-in time after a user has finished his changes. To be able to catch conflicts while developers are implementing their tasks, workspace awareness techniques were proposed. Tools such as IBM's Jazz.net platform \cite{ref_Jazz} or Microsoft's CollabVS system \cite{ref_CollabVS} augment the awareness of developers and propagate changes at file/class level immediately after they happen. Also, some researchers have investigated how to exploit the information produced by integrated development environments during development such as Mylyn \cite{ref_Mylyn}, Sypware \cite{ref_Spyware}, and Syde \cite{ref_synchronus_development_Hattori}. Palantir \cite{ref_Indirect_Conflict_Awareness_Sarma} is the first awareness tool that tries to detect indirect conflicts in addition to direct conflicts. A new approach to detect potential indirect conflicts was presented in CASI \cite{ref_CASI}.

However, none of the previous works mentions conflicts among UML model elements. AMOR \cite{ref_AMOR}, SMoVer \cite{ref_SMoVer}, and COMOVER \cite{ref_COMOVER} are model versioning systems that consider conflicts among model elements. Nevertheless, like other VCSs, conflicts in these systems are just detected at check-in time. Also, conflict caused by concurrent changes is just one source of inconsistency.

ADAMS \cite{ref_ADAMS} is an example of a different approach to support distributed collaborative work. It is a web-based system that integrates project
management features and artifact management features. 

\section{Conclusion}
In this paper, we have presented a Change Support Model for distributed collaborative work. To help change workers implement change activity safety and efficiently, CSM generates Change Support Workflows based on dependency relationships among software artifacts. Towards CSW generation, we have proposed a model and a tool for automatic generation of Basic Dependency Relationships among UML model elements. In additon, CSM also considers inconsistency caused by concurrent CSWs that work on shared artifacts or dependent artifacts. To detect inconsistencies as soon as possible, our method detects potential inconsistencies at both build time and runtime. We have also identified inconsistency patterns to help detect inconsistencies at runtime more effectively. CSM will help collect and analyze information of local CSWs to notify change worker about risky points. Some inconsistency resolutions have been proposed too. 
%Finally, we have introduced an Impact Analysis tool that generates BDRs automatically. 

%As future work, we will improve our method on inconsistency handling, especially inconsistency analysis and resolution. Next, we will work on generating dependency relationships connecting UML model elements to source code. Finally, we will develop a complete prototype of CSS that includes CSW generation and management.
In future work, we will improve our method of inconsistency handling, especially inconsistency analysis and resolution. Next, we will work on generating dependency relationships connecting UML model elements to source code. Finally, we will develop a tool that implements CSW generation and inconsistency awareness support.



%\bibliographystyle{IEEEtran}

\end{document}